\documentclass[12pt]{article}

\usepackage{amsfonts, amsmath}

\newcommand{\be}{\begin{equation}} \newcommand{\ee}{\end{equation}}
\newcommand{\lbl}[1]{\label{#1}}
\newcommand{\bH}[1]{\ensuremath{\hat{\boldsymbol{#1}}}}
\newcommand{\scal}[2]{\langle #1\left|#2\rangle\right.}
\newcommand{\vlmd}{\ensuremath{\boldsymbol{\lambda}}}
\newcommand{\vvphi}{\ensuremath{\boldsymbol{\varphi}}}

\def\cU{\ensuremath{U(\mathcal{C})}}
\def\A{\ensuremath{\boldsymbol{A}}}

\def\ndd{\noindent}
\def\bsm{\boldsymbol}

\begin{document}
\thispagestyle{empty}

\title{Geometric Phases and Quantum Computations}
\author{A.E.Margolin\hspace{1.5mm}$^*$, V.I.Strazhev
\thanks{National Centre of High Energy Physics,
                         Bogdanovich Str.153, Minsk 220040, Belarus}
\hspace{1mm} and \hspace{1mm} A.Ya.Tregubovich
\thanks{Institute of Physics National Academy of Sciences
                          Skoryna av.68, Minsk 220072, Belarus}}

\date{}
\maketitle

\begin{abstract}
Calculation aspects of holonomic quantum computer (HQC) are
considered. Wilczek--Zee potential defining the set of quantum
calculations for HQC is explicitly evaluated.  Principal
possibility of realization of the logical gates for this case is
discussed.
\end{abstract}

 The conception of quantum computer (QC) and quantum calculation developed in
80-th \cite{Deutsch1}--\cite{Feynman} were found to be fruitful both for
computer science and mathematics as well as for physics \cite{Steane}. Although
a device being able to perform quantum calculations is far away from practical
realization, there is a number of theoretical proposals of such a construct
\cite{Lloyd1}--\cite{Gershenfeld}. Recently holonomic realization of QC was
proposed \cite{Zanardi1},\cite{Zanardi2}. It is based on the notion of
non-Abelian Berry phase\cite{Wilczek-Zee}.To perform quantum computations in
this approach one needs a parametrically driven quantum system described by the
Hamiltonian $H(\vlmd ) $ with adiabatically evolving parameters $\vlmd (t) =
(\lambda_1(t),\ldots
 ,\lambda_N(t))$. Adiabaticity means that $\Omega \ll \omega_{min}$
 where $\Omega$ is the characteristic frequency in the Fourier spectrum of
 $\vlmd (t)$ and $\omega_{min}$ is the minimal transition frequency in the
 spectrum of $H(\vlmd )$. If the spectrum is degenerate then the cyclic evolution
 of $N$ parameters $\lambda_A(t)$ results in a unitary transformation of
 each eigenspace of $H$. Such transformations can be treated as computations
on the eigenspace representing a part of the qubit space of HQC. These
computations are shown to form a universal set of quantum gates for a
specific relevant model \cite{Pachos1}. For instance, one should choose
of independent 2-loops $\mathcal{C}$ in the control manifold $ M =
\{\text{all possible}\, \vlmd \}$
 and then the non-Abelian Berry phases
\cU\ corresponding to each loop represent all basic gates. To produce
such gates one should have a possibility to control i.e. for each closed
adiabatic curve $\mathcal{C}$ in $M$ one is to know the explicit form of
\cU. To know that it is necessary to be aware of an explicit expression
of non-Abelian Wilczek--Zee potential \A\ which determines \cU\
completely. A more or less universal method of explicit calculation of
\A\ is to our knowledge up to now absent. For an early overview on this
subject see \cite{Wilczek-Shapere}. A variant of Berry phase evaluation
for a finite level system relevant to HQC was proposed in \cite{Zanardi2}
where a specific parametrization of $SU(n)$ is used. In this paper we
present some less general but to our mind some more effective method of
evaluation of non-Abelian Berry phase which can appear to be useful for
HQC. For this purpose we restrict ourselves with systems with finite number energy
levels described by isospectral
Hamiltonians, i.e. by ones whose spectrum does not depend on time via
adiabatical evolution of their parameters.
 At first we demonstrate the method for the Abelian case, then derive
the result for the non-Abelian one and in conclusion  some possibilities of physical
realization of the basic gates.\par
In veew of above made restrictions the Hamiltonian we consider can be present in
the form

\be\lbl{starthamilt}
     H(t) = V(t)H_0V^{\dagger}(t),
\end{equation}

\ndd where the $n\times n$ matrix $H_0$ does not depend on time. As the qubit
space must be large enough then the state space of $H$ must be so as well. In
this case derivation of \A\ becomes difficult even if there are few
independently evolving parameters that is the case for a typical experimental
situation.

We derive an explicit closed expression for \A\ in terms of $H_{ij}(t)$ that
have a direct physical sense. Let us consider the Hamiltonian $H(\vlmd )$ with
$\lambda_A(t)$ evolving adiabatically thus Born--Fock theorem \cite{Messiah}
 is valid:

\be\lbl{AdiabSchroed}
H(\vlmd ) \varphi_i^{a_i}(\vlmd ) = E_i(\vlmd ) \varphi_i^{a_i}(\vlmd ),
\qquad   i=1,\ldots ,s, \quad a_i=1, \ldots d_i.
\end{equation}

\ndd Here $d_i$ is the degeneration rate and $\sum_{i=1}^{s}d_i = n$.
It should be emphasized that the finite-matrix form of the Hamiltonian $H(\vlmd)$
may be not only the result of an approximation, i.e. one neglects all energy
levels whose number is greater than $n$ but also a consequence of the symmetry
of the system under consideration. This means if the Hamiltonian $\bH{H} $ we
start with possesses a Liean symmetry, i.e. it can be represented in the form
$\bH{H} = c_A(\vlmd )\,\bH{T}_A$, where $\bH{T}_A$ are generators of a Lie
algebra, then the state space of the system is split in a set of components
irreducible with respect to the algebra and for each finite-dimensional
component we come to the finite-matrix form.

Cyclic evolution of \vlmd\ results
\cite{Wilczek-Zee} in a monodromy transformation of
$\vvphi = (\varphi_i^1,\ldots ,\varphi_i^{d_i})$:
\be\lbl{defnonAbPhase}
\vvphi (T) = \cU \vvphi (0),
\end{equation}

\ndd where $t$ parametrizes the curve $\mathcal{C}$ so that $\vlmd (T)=\vlmd (0)$.

\ndd For  an illustration we start with Abelian Berry phase that was
firstly predicted by Berry \cite{Berry}. In this case all $d_i=1$ and instead of
(\ref{defnonAbPhase}) the relation
 \be\lbl{defAbPhase}
 \varphi_m (T) = e^{i\gamma_m (\mathcal{C})} \varphi_m (0),
 \end{equation}
where
\be\lbl{defPotAb}
\gamma_m (\mathcal{C}) = \int\limits_{\mathcal{C}} A_m\, ,\qquad
A_m = i\, \scal{\varphi_m}{d\varphi_m}.
\end{equation}
As the spectrum of the Hamiltonian is finite, the ket
$\left.\left|\varphi_m\right>\right.$ is a unit vector $\bsm{m}$ in
$\boldsymbol{\mathrm{C}}^n$, so $A_m$ is
\be\lbl{abAn1}
A_m = {i\over 2}\, (\bsm{m}^*d\bsm{m} - \bsm{m}d\bsm{m}^*).
\end{equation}
As the evolution is adiabatic, the spectrum of $H(t)$ remains always nondegenerate
if it was so at the initial time. Then there is always a nonzero main minor of
$H - E_m$ which we assume to consist always of the first $n-1$ lines and columns of
$H - E_m$. Denoting the matrix consisting of the first $n-1$ lines and columns
of $H$ by $H_{\perp}$ we come to the condition
\be\lbl{condition}
\det (H_{\perp} - E_m) \neq 0
\end{equation}
Making use of this condition one can represent $\bsm{n}$ in tyhe uniform
coordinates
$$ \bsm{m} = \frac{(\bsm{\xi}_m, 1)}{\sqrt{1+|\bsm{\xi}_m|^2}}  $$
and express $\bsm{\xi_m}$ in terms of $H_{ij}$ for $1\le i,j\le n-1$ and $E_m$:
\be\lbl{xin}
\bsm{\xi}_m = (H_{\perp} - E_m)^{-1}\, \bsm{h}, \qquad h_i = -H_{in},
\end{equation}
 where $\bsm{h}$ is a vector in $\boldsymbol{\mathrm{C}}^{n-1}$ but not in
 $\boldsymbol{\mathrm{C}}^n$. Thus we have for $A_m$
 \be\lbl{abAn2}
A_m = {i\over 2}\, \frac{(\bsm{\xi_m}^*d\bsm{\xi_m} - \bsm{\xi_m}d\bsm{\xi_m}^*).}
 {1+|\bsm{\xi}_m|^2} ,
 \end{equation}
 where $\bsm{\xi}_m$ is completely determined by (\ref{xin}). Note that the result
 obtained is purely geometrical because it can be expressed of the K{\"a}hlerian
 potential
 $$  F = \log (1 + |\bsm{z}|^2) $$
 which determines all the geometrical properties of the state space
 $$ SU(n)/\underbrace{U(1)\times ...\times U(1)}_{n-1\;\mathrm{times}}. $$
Here $\bsm{z}$ is a vector in $\boldsymbol{\mathrm{C}}^{N^2}$ consisting of
$n(n-1)/2$ independent components of all $\bsm{\xi}_m$. One more observation is that
for the case $n=2$ corresponding to spin in the magnetic field (\ref{abAn2}) gives
well known result \cite{Berry}
$$    A_{\pm} = \pm{1\over 2}\, \oint\limits_{\mathcal{C}}
                 \frac{\xi^*d\xi- \xi d\xi^*}{1 + |\xi |^2} =
                 \pm{1\over 2}\, \Omega (\mathcal{C}),
$$
where $\pm$ labels spim-up and spin-down states, $\xi$ is thought of as a
stereographic projection of a point on the unit sphere
$S^2$ and $\Omega (\mathcal{C})$ is the solid angle corresponding to the given closed contour
 on the sphere.

\par Now we proceed with a more general case of degenerate spectrum. Quantum computation
for this case generated by an adiabatic loop in the control manifold is determined
 by (\ref{defnonAbPhase}) where $\cU$ is presented by a $\mathcal{P}$-ordered
 exponent
 \be\lbl{nonAbPhase}
 \cU = \mathcal{P}\exp \left( \oint_{\mathcal{C}}\A_m\right),\qquad
 (\A_m)_{ab}=i\scal{m_b}{dm_a}.
 \end{equation}
 The set of
eigenvectors $\bsm{\xi}_{ma}$, $a=1,...,d_m$ must obey the equation
\be\lbl{xina}
(H_{\perp}^{(d_m)} - E_m)\,\bsm{\xi}_{ma} =  h\, \bsm{c}_a.
\end{equation}
Here the matrix $H_{\perp}^{(d_m)}$ is constructed from the first $n-d_m$ lines
and columns of $H$, $\bsm{c}_a$ are arbitrary $d_m$-dimensional vectors and $h$ is
the following $(n-d_m)\times d_m$-matrix:
$$  h = - \begin{pmatrix} H_{1, n-d_m+1} & \ldots  & H_{1,n}\\
                           \vdots      & \ldots  & ...     \\
                    H_{n-d_m, n-d_m+1} & \ldots  & H_{n-d_m,n}  \end{pmatrix}
$$
Of course it has sense only if the condition
\be\lbl{cond}
\det (H_{\perp}^{(d_m)}(t) - E_m) \neq 0
\end{equation}
is valid along the evolution process. The set of vectors $\bsm{\xi}_{ma}$ must
be orthogonalized by the standard Gram algorithm and after that we get the
orthonormal set of the eigenvectors $\bsm{z}_a$ (here and below we has omitted
the index $m$) in the form
\be\lbl{orthset}
\bsm{z}_a = {1\over \det\Gamma_{a-1}}\,
\begin{pmatrix}
                                 &              &       & \bsm{x}_1   \\
                                 & \Gamma_{a-1} &       & \vdots         \\
\scal{\bsm{\xi}_a}{\bsm{\xi}_1}  & \ldots       &
                      \scal{\bsm{\xi}_a}{\bsm{\xi}_{a-1}} & \bsm{x}_a
\end{pmatrix} ,
\end{equation}
where $\bsm{x}_b=(\bsm{\xi}_b, \bsm{c}_b)$ and $\bsm{c}_a$ is chosen to be the
standard orthogonal set $\bsm{c}_a=(0...\overbrace{1}^a ...0)$.
The matrices $\Gamma_a$ are determined by
\be\lbl{Gamma}
\Gamma_a = \begin{pmatrix}
1+\scal{\bsm{\xi}_1}{\bsm{\xi}_1}&\ldots  & \scal{\bsm{\xi}_1}{\bsm{\xi}_a} \\
     \vdots                      & \ddots & \vdots                           \\
\scal{\bsm{\xi}_a}{\bsm{\xi}_1}  &\ldots  &1+\scal{\bsm{\xi}_a}{\bsm{\xi}_a}
\end{pmatrix} = 1 + Z_a^{\dagger}Z_a ,
\end{equation}
where the $(n-d_m)\times a$-matrix $Z_a$ consists of $a$ first lines of the
$(n-d_m)\times d_m$-matrix $Z = (H_{\perp}^{(d_m)} - E_m)^{-1}\, h$.
Using (\ref{Gamma}) and (\ref{orthset}) we come to the final expression for the
matrix-valued $1$-form \A :
\be\lbl{finalA}
\A = {i\over 2}\;
\frac{
g_{ab}^{ij} (\bsm{\xi_j}^*d\bsm{\xi_i} - d\bsm{\xi_j}^*\bsm{\xi_i}) + 2\,\omega_{ab}
}
{  \det ( 1 + Z_{a-1}^{\dagger}Z_{a-1})\, \det ( 1 + Z_{b-1}^{\dagger}Z_{b-`})  },
\quad 1\le i \le a,\; 1\le j\le b,
\end{equation}
where
$$
 g_{ab}^{ij} =  \Gamma_a ^i\Gamma_b ^{*j}, \qquad
 \omega _{ab} = \scal{\bsm{\xi}_j}{\left. d\,\mathrm{Im}(g_{ab}^{ij})
                                                  \right|\bsm{\xi}_i}
 + \sum\limits_{i=1}^{min(a,b)}d\,\mathrm{Im}(g_{ab}^{ii}),
$$
and $\Gamma_a^i$ is the cofactor of $\bsm{\xi}_i$ in $\Gamma_a$. Note that the
change of our basis $\bsm{c}_a$ by $\bsm{c}_{a}^{\,\prime} = U_{ab}(\vlmd )\, \bsm{c}_b$
leads to a standard gauge transformation of \A\:
$$
 \A^{\prime} = U\A U^{\dagger} + i(dU)U^{\dagger}
$$
The formula
(\ref{finalA}) is the desired expression of \A\ in terms of the matrix elements of the
Hamiltonian. It is correct if condition (\ref{cond}) is valid. It is not nevertheless
a principal restriction because $d_m$ does not depend on time due to adiabaticity
of the evolution and there is always at least one nonzero $n-d_m$-order minor of
$H$. Then, if the minor we choose vanishes somewhere on the loop $\mathcal{C}$ one
can always take local coordinates such that the techniques considered is applicable
on each segment of $\mathcal{C}$.

One more thing to discuss here are quantum gates and their possible physical
realization. As the Hamiltonian is an element of a Lie algebra which we assume
to be semisimple, generic computations are of course elements of the corresponding
Lie group. We represent them in the form
\be\lbl{genElGr}
U = \exp \left(\sum\limits_{i=1}^k y_i H_i\right)\,
    \prod\limits_{\alpha}\exp \left(\zeta_{\alpha}E_{\alpha} -
                                \zeta_{\alpha}^*E_{-\alpha} \right),
\end{equation}
where $y_i$ are real and $\zeta_{\alpha}$ are complex parameters, $k$ is the
dimension of the Cartan subalgebra,  $i$ labels all it's linearly independent elements,
 the product is taking over all positive roots
of the algebra, and the standard notations for Cartan--Weyl basis have been used.
As the action of the first exponent in (\ref{genElGr}) results in a trivial phase
factor, a generic computation is determined by the product over all roots. Thus we come
to the conclusion that the elementary factors
 \be\lbl{gate}
 U_{\alpha} = \exp \left(\zeta_{\alpha}E_{\alpha} -
                                 \zeta_{\alpha}^*E_{-\alpha} \right),
 \end{equation}
provide the basis of quantum gates for the model. Their number is e.g. for the
first fundamental series $A_n$ $n(n-1)/2$ as it should be. An idea of physical
realization of the gates appears if one takes into account that for $A_n$
$E_{\alpha}$ can be realized by means of ordinary bosonic creation and annihilation
operators, namely $E_{\alpha}=a_i{\dagger}a_j$ for some $1\le i,j \le n$. Then
$E_{\alpha}$ represents nothing but two-mode squeezing operator. Thus the model
considered can be applied to optical HQC with $n$ laser beams (the case n=2 is
considered in \cite{Pachos1}) and the logical
gates $U_{\alpha} $ are just two-qubit transformations realized by transformation
of two laser beams.

The method presented here enables one to build in principal any computation for
HQC described by a Hamiltonian with a stationary spectrum in terms of experimentally
measured values, namely the matrix elements of the Hamiltonian. The method depends
weakly on the dimension of the qubit space which other models based on various
parametrizations of evolution operators of the system are very sensitive to.
Application of this method to a concrete physical model will be discussed
elsewhere.

\end{document}